# Permeability of a bubble assembly:
# From the very dry to the wet limit

Florence Rouyer, Olivier Pitois, Elise Lorenceau, Nicolas Louvet

Université Paris-Est, Laboratoire de Physique des Matériaux Divisés et des Interfaces,

UMR CNRS 8108. 5 bvd Descartes, 77454 Marne la Vallée Cedex 2, France.

**Abstract**

We measure the permeability of a fluidized bed of monodispersed bubbles with soap solution characteristic of mobile and non-mobile interfaces. These experimental data extend the permeability curves previously published for foam in the dry limit. In the wet limit, these data join the permeability curves of a hard sphere suspension at porosity equal to 0.4 and 0.6 in the cases of mobile and non-mobile interfaces respectively. We show that the model of permeability proposed by Kozeny and Carman and originally validated for packed beds of spheres – with porosity around 0.4 – can be successfully applied with no adjustable parameters to liquid fractions from 0.001 up to 0.85 for systems made of monodisperse and deformable entities with non-mobile interfaces.

## I. Introduction:

What is common to foam drainage, flow in porous media made of packed beads and settling suspensions? From a physical point of view, for all these cases, these systems are permeable and viscous liquid flows through interstices of the assembly of individual entities (bubbles, drops or beads). In these systems, the Reynolds number is generally low meaning that viscous effects are dominant compared to inertia. However many differences can be listed : interfaces are fluid in a foam, as opposed to solid interfaces in porous media and solid suspension; the size of the interstices varies over a wide range (0.01mm-10mm); the interstices can either be static or in constant evolution in the fluidized regime due to hydrodynamic agitation of the individual entities. Note that these phenomena are encountered in many engineering and biomedical applications and are all very active research areas. For all these reasons, these particulate assemblies have generally been studied independently.

The parameter commonly used to characterize the ability of liquid to flow through a media is the hydrodynamic permeability [1]. The permeability of porous material with porosity ($\varepsilon$=liquid volume fraction) around 0.4 is well described by the Carman-Kozeny law [2, 3] and many experiments and simulations have explored the robustness of this law [4]. To our knowledge, few works have tested this law in the limit of very dry ($\varepsilon<0.05$) [5] or very wet ($\varepsilon>0.6$) media [6]. In case of foams that have fluid interfaces, permeability values as well as variation of permeability with porosity depend on the mobility of the interfaces [7, 8, 9]. In particular, permeability values are larger for mobile interfaces than for non-mobile interfaces.

In the present work, we provide new experimental data for the permeability of a fluidized bed of monodisperse bubbles with distinct interface mobility to cover the intermediate range of porosity that connects the dry and wet limits, respectively achieved for foam and suspension of hard-spheres. We first give a brief overview of foam and suspension permeability. Next, we present the experiment of a stable fluidized bed made of bubbles and their results for the permeability versus porosity for two extreme kinds of interface mobility. For the case of non-mobile interfaces, we eventually test the Kozeny-Carman law to describe the permeability of a bubble assembly over 3 decades of the liquid volume fraction.

## II. Permeability of foam and suspension:

In 1856, Darcy formulated an empirical equation based on measurements of the flow of water through sands and sand-stones that relates the superficial liquid velocity (apparent velocity) to the pressure gradient responsible for the flow [1]. The Darcy law is commonly written as follows:

$$u = \frac{k}{\mu}\frac{\Delta P}{L}$$

where $\mu$ is the bulk dynamic shear viscosity of the flowing fluid and $k$ is the permeability. $k$ quantitatively describes the ability of the solid porous material to let fluids flow through it. The dimension of $k$ is the square of a length. In the following, we normalize the permeability $k$ by D² where D is the diameter of an individual entity, meaning diameter of the spherical bubble.

### A. Low liquid fractions- Foam permeability:

In the limit of low liquid fractions, the monodisperse particulate system is made of soft deformable entities that are packed together. Foams (or emulsions) made of bubbles (droplets) are such media. The permeability of foam is usually modeled by considering the liquid phase which is in the network of Plateau borders (films junctions) and nodes (Plateau border junctions). Indeed, flows in films are neglected due to the low permeability of films and their relatively small volume compared to the whole liquid volume contained in the foam. Two primary models are commonly used to interpret foam drainage experiments. Weaire et al. proposed a model for dry foams [7], where the viscous dissipation is located in the network of long and slender Plateau border channels - assuming a Poiseuille flow in the channels and non-mobile interfaces. According to this first model, the foam permeability is given by $k_c = 8.10^{-4} D^2 \varepsilon^2$. It describes the channel dominated regime and is classically associated to foams made of non-mobile interfaces that correspond to the limit case of infinite surface shear viscosity – zero velocity at the interface which acts like a solid wall. Considering mobile interfaces, Koehler et al. [8] have located the viscous dissipation in nodes. This yields for the foam permeability $k_c = 6.7.10^{-4} D^2 \varepsilon^{3/2}$. This later model describes the node dominated regime and is associated to foams made of mobile interfaces that correspond to the limit case of zero surface shear viscosity – slipping wall. Depending on the foaming agents used, these two models have been extensively validated in the limit of relatively dry foams ($\varepsilon<0.07$). Yet, at higher liquid fractions, there is less agreement between the experimental data and the scaling laws; the permeability of foam can not be fitted by a unique power law over a large range of liquid fraction [9]. This can be observed in figure 1, where typical data for foam permeability are displayed in full symbols for bubble size ranging from 0.7 to 2.2mm.

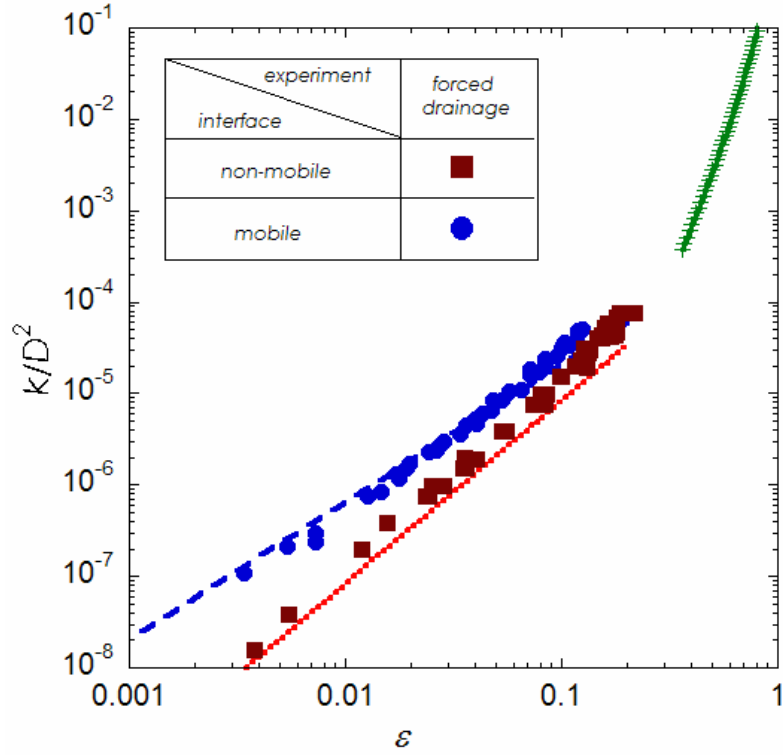

**Figure 1**: Permeability of foams made of monodisperse bubbles (see legend for details) as a function of porosity. Models detailed in the text are represented by the red dotted line [7] and the dashed blue line [8]. The Richardson-Zaki law (RZ law) is represented by the green plus sign curve.

### B. High liquid fractions - Suspension permeability models:

In the limit of high liquid fractions, the monodisperse particulate system is made of disjoined entities. One of the simplest system that comes to mind is a suspension of non-Brownian hard spheres. Under gravity, buoyant particles settle and due to back flow effect, the settling velocity $V_{sed}$ - equal to the average velocity of the particles - decreases as a function of solid volume fraction because the viscous friction increases as the size of the interstices between particles decreases. In 1954, Richardson and Zaki presented an empirical law (RZ) to fit the dependence of the average settling velocity as a function of solid volume fraction ($\phi=1-\varepsilon$): $V_{sed}= V_{Stokes}(1-\phi)^{5.5}=V_{Stokes}\varepsilon^{5.5}$ where $V_{Stokes}$ is the settling velocity of an isolated sphere in an infinite medium [10]. Up to now, this empirical law is the reference for experimental average settling velocity of suspensions [11], and it is also valid to describe the upward fluid velocity in case of fluidized suspensions [12]. In 1994, Mills and Snabre used a mean field approach to estimate the energy dissipation during the homogeneous sedimentation or the particulate fluidization of non-Brownian hard spheres in a concentrated suspension that allows one to retrieve the RZ fit in the limit of low Reynolds number (Re), where inertial effects are negligible[13]. They model the viscous dissipation by considering the average distance between particles. This typical distance can be evaluated considering the specific surface area of the particles. However to satisfy to Batchelor theory [14], they need to adjust a constant in the dilute regime. In later work, they add to their model inertial effects and conclude that the average settling velocity of suspensions can be described by the law $V_{sed} = V_{Stokes}(1-\phi)^n$ where n is a decreasing function of the particulate Reynolds number ($Re_p$) [13].

Permeability of suspensions of non-Brownian particles can be deduced from sedimenting and from fluidizing velocities. In case of sedimentation, the suspension is at rest i.e. $\langle v_{susp} \rangle = \varepsilon \langle v_f \rangle + (1-\varepsilon) \langle v_p \rangle = 0$ and the apparent velocity of the fluid relative to the particles is equal to $\varepsilon(\langle v_f \rangle - \langle v_p \rangle) = -\langle v_p \rangle = V_{sed}$. $v_{susp}$, $v_f$, $v_p$ are respectively the local velocity of the suspension, of the fluid and of the particles; $\langle \ \rangle$ denotes average over space and/or time; $V_{sed}$ is the front velocity of the sedimenting suspension. One can impose the particles to be at rest in average in the frame of the laboratory by imposing to the fluid an upward flow rate Q. This is called fluidization. In that latter case, the average velocity of the particles is equal to zero and thus the apparent fluid velocity relative to the particles is equal to $\varepsilon \langle v_f \rangle = Q/A = u$ where $A$ is the cross-section of the sample. The permeability of the suspension is deduced from the Darcy law with a pressure gradient induced by buoyancy (difference of the gravity forces that act on the suspension and on the fluid) [15, 16]:

$$k_{susp} = \frac{\mu V}{\Delta \rho g (1-\varepsilon)} = \frac{1}{18} \frac{V}{V_{Stokes}(1-\varepsilon)} D^2 \quad \text{(equation 1)}$$

where $V$ is either equal to $V_{sed}$ or $u$.

We present in figure 1 the permeability of a suspension calculated from the above equation and considering the Richarson-Zaki law $V_{sed} = V_{Stokes}\varepsilon^{5.5}$ for $\varepsilon > 0.36$. On this log-log plot, the experimental data of foam drainage and models currently used in the limit of low and high porosity, are shown as well. It can be observed that these two sets of data are monotonically increasing, yet they do not overlap and there is a gap of liquid fraction $0.2 < \varepsilon < 0.36$ for which no description is available. In the following, we propose an extension of the Carman-Kozeny model – originally validated for packed bed spheres ($\varepsilon \approx 0.46$) to describe this range of liquid fraction.

### C. Intermediate liquid fractions - modified Carman-Kozeny models:

Considering the pressure drop within the pore of a porous medium, Kozeny writes $u = \frac{1}{\mu} \frac{\varepsilon^3}{cS^2} \frac{\Delta P}{L}$, where $\varepsilon$ is the porosity (liquid volume fraction) and $S$ the solid surface per volume also named specific surface area and $c$ is a dimensionless parameter [2]. Physically, one can represent the specific surface area as the inverse of the characteristic length of channel that constitutes the porous medium (mean hydraulic radius), and thus the length scale that has to be taken into account for estimating viscous dissipation via the velocity gradient. One can write $S=\varepsilon/m$, where $m$ is mean hydraulic radius. For porous media made of a hard sphere packing, it is straightword to determined the specific surface area that is equal to the number of particles per unit volume ($n$) times the surface of each particles $S = n(\pi D^2) = 6(1-\varepsilon)/D$. In 1937, it was demonstrated that Carman's experiments verified Kozeny equation with $c=5$ for a granular bed made of monodispersed hard spheres [3]. Since these pioneering works, the permeability of media is referred as the Carman-Kozeny law:

$$k = \frac{\varepsilon^3}{5S^2} \text{ (equation 2).}$$

This law that only requires the knowledge of porosity and specific surface is of great interest for describing flow in porous media with complex geometry. This equation is all the more valid that the porous media is made of random packing of spheres [17].

When considering deformable entities such as bubbles of an aqueous foam, the determination of specific surface is less obvious than for hard spheres packing. For monodisperse bubbles, if the continuous phase filled, for example, less than 32% of the total volume for a bcc structure (Kelvin cell structure, [18]), the bubbles are not spherical and each bubble shares a film (two very close interfaces) with its neighbors. As the films do not contribute to liquid drainage, neither of their surface areas contribute to $S$. To determine the geometry and consequently the specific surface area of such porous media, we use the equilibrium state of the foam that corresponds to the minimal interface area (note that the energy of this system is equal to surface tension of the liquid/air interface times the total interface surface). The variation of the specific surface of foam for $\varepsilon$ varying from zero up to 0.32 is thus deduced from previous works with the Surface Evolver software [19] that has computed the surface of films [20] and surface of bubbles [21] as a function of liquid fraction [22]. As for porous media made of hard spheres, the normalized specific surface area of a suspension of undeformed spheres (bubbles or solid particles) is equal to $SD = 6(1-\varepsilon)$. Figure 2 represents the variation of the normalized specific surface area as a function of $\varepsilon$ for both foam up to 0.32 and suspension for $\varepsilon$ from 0.26 (maximal close packing of hard spheres) up to 1. At small porosity ($\varepsilon < 0.05$), $S$ varies as the square root of $\varepsilon$: $S(\varepsilon)D \approx 15.8\sqrt{\varepsilon}$ (see inset figure 4). The maximal value of the specific surface area is reached for porosity around 0.2. Note that a continuous function can be drawn. However the first derivative of this function is not continuous. Indeed, near the close packing limit, the specific surface area is highly sensitive to the packing structure (spatial arrangement of the spheres, i.e. bcc, fcc, random …). To cover the whole range of liquid fraction (i.e. from 0 to 1), we introduce a function $g(\varepsilon)$ that satisfies the limit conditions $g(0)=1$ and insures the continuity of the function $S(\varepsilon)$ for $\varepsilon = 0.4$, i.e. $S(0.4)D=3.6$ and $\frac{dS}{d\varepsilon}(0.4)D = -6$. The specific surface is thus describes as $S(\varepsilon)D \approx 15.8\sqrt{\varepsilon}\, g(\varepsilon)$ for $\varepsilon <$ 0.4 and $S(\varepsilon)D = 6(1-\varepsilon)$ for $\varepsilon > 0.4$. We note that this function $g(\varepsilon)$ is not a fit but rather a basic function that permits us to connect the description at low and high porosity. Assuming that $g$ is a function of second order, we find $g(\varepsilon)=1-2.15\varepsilon+1.37\varepsilon^2$. As discussed in the next section, the variation of the specific surface around $\varepsilon=0.3$ has little influence on the description of the permeability.

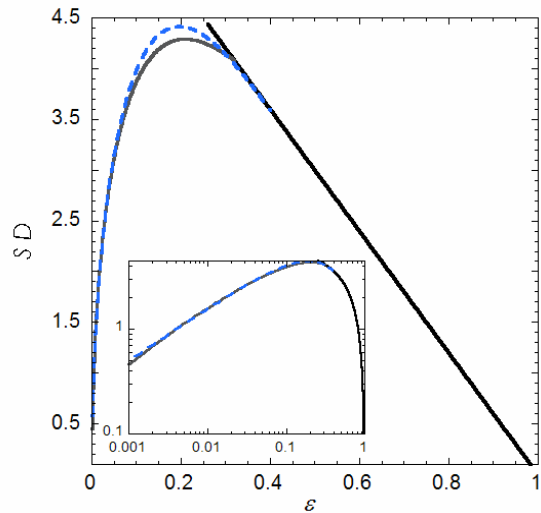

**Figure 2**: Variation of the normalized specific surface area of particulate systems made of monodisperse entities as a function of porosity. For $\varepsilon <0.32$ (grey) the specific surface is estimated from Surface Evolver simulation as explained in the text. For $\varepsilon > 0.36$ (black) $SD = \pi D^3 n = 6(1-\varepsilon)$. The blue dotted line corresponds to the function $S(\varepsilon)D \approx 15.8\sqrt{\varepsilon}\left(1-2.15\varepsilon+1.37\varepsilon^2\right)$. Inset : log-log plot.

Once the specific surface area is estimated for monodisperse foams or perfectly known for suspension, the permeability of these media can be estimated according to the Carman-Kozeny law (equation (1)) with $c=5$, both for aqueous foam ($\varepsilon< 0.4$) and for suspension ($\varepsilon> 0.4$). It yields:

$$k(\varepsilon) = \frac{\varepsilon^2}{1248\left(1-2.15\varepsilon+1.37\varepsilon^2\right)^2} D^2 \text{ for } \varepsilon<0.4 \text{ (equation 3.1)}$$

$$k(\varepsilon) = \frac{\varepsilon^3}{180(1-\varepsilon)^2} D^2 \text{ for } \varepsilon>0.4 \text{ (equation 3.2)}$$

To validate this approach, we carry out new experiments and measure the permeability of fluidized bed of monodisperse bubbles in the range of liquid fraction where no data are available.

## III. Fluidized bed of monodisperse bubbles :

To gain better understanding on the influence of the boundary conditions of the media, we use two different soap solutions either providing rigid walls - no slip boundary conditions- or mobile walls. This new set of data connects the two limits of low and high porosity and we compare it with the modified Carman-Kozeny model.

### A. Experimental Set-up and protocol :

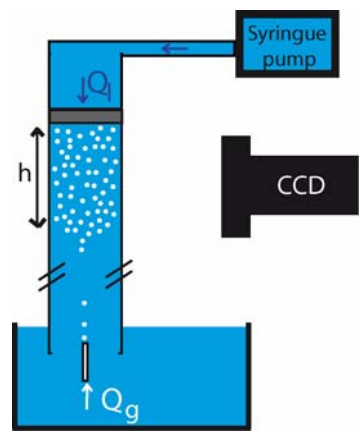

**Figure 3:** Sketch of the experimental setup of the fluidized bed.

To obtain a fluidized bed of bubbles, a Plexiglas column is set vertically and partially immersed in a soap solution bath. The top of the column ends with a grid (grey on figure 3) and is connected to a syringe pump that delivers soap solution at fixed liquid flow rate $Q_l$ in the range of 0.01-10 ml/min. A needle is placed at the open bottom of the column and allows monodisperse bubbles to be produced at constant gas flow rate $Q_g$ (cf. figure 3). The mesh size of the grid is smaller than the bubble diameter that is of the order of hundred micrometers. Gas is saturated with perfluorohexane to avoid size evolution of the bubble at the time scale of the experiment (hour). The radius and height of the column are respectively equal to 4.75mm and 200mm. Similarly to the experiments that measured the permeability of foam [9], two soap solutions are used, which are known for their different interface mobility (mobile or non-mobile interfaces). The mobile interface is obtained for an aqueous solution made of 3g/L of TTAB (TetradecylTrimethylAmmoniumBromide). The non-mobile interface is obtained for an aqueous solution made of 3g/L of TTAB and 0,2 g/L dodecanol. For both solutions, surfactant concentration is well above the CMC (1g/L), the density is $\rho$ = 1000 kg/m$^3$, the bulk shear viscosity is $\eta$ = 1 mPa.s. We report in the next table the surface tension

$\gamma$, the surface shear viscosity $\eta_s$ [23] and the bubble diameter $D$ for each soap solution. The particulate Reynolds number $Re_p$ is of the order of 4.

| Interface | $\gamma$ (mN/m) | $\eta_s$ (g.s$^{-1}$) | D (µm) |
|---|---|---|---|
| Mobile | 38 | ~$10^{-5}$ | 260 |
| Non-mobile | 25 | ~$10^{-3}$ | 190 |

Initially, a small liquid flow rate ($Q_l$ < 1 ml/min) is imposed from top to bottom of the cylinder while monodisperse gas bubbles are produced with constant gas rate ($Q_g$) at the bottom of the column with a needle, cf. figure 3. The particulate system made of bubble grows at the front velocity dh/dt, its evolution is recorded by a CCD camera. We checked that dh/dt is constant during the bed production. Then, the gas flow rate is stopped when the height of the cluster of bubbles h is larger than at least three column diameters, this initial height is noted $h_i$. Next, the bubble bed is fully fluidized by a large liquid flow rate ($Q_l$>5 ml/min). Finally, different flow rates are successively imposed and the height h of the bubble bed is measured. The initial porosity $\varepsilon_i$ and the porosity of the bubble bed for each liquid flow rate are respectively calculated by the following equations: $\varepsilon_i = \dfrac{Q_g}{(\pi R^2)\dfrac{dh}{dt}}$ and $\varepsilon = 1 - (1-\varepsilon_i)\dfrac{h_i}{h}$.

The permeability is obtained from equation (1) where V is equal to Ql/$\pi R^2$.

**B. Results :**

Normalized permeability curves are shown in figure 4 as a function of porosity for the two soap solutions. The permeability data measured from foam drainage and fluidization of bed of bubbles of the two different soap solutions overlap for porosity range around 0.1-0.2. The data of fluidized bed with non-mobile interface bubble superimpose data of the permeability of a suspension made of solid particles (RZ law represented by the green line) for $\varepsilon$ > 0.36). As expected, the permeability data of fluidized bed with mobile interface are larger than in the case of non-mobile interface (less viscous dissipation); however for large porosity $\varepsilon$ > 0.55, these data tend to overlap with the data obtained with solid particles.

The data of the non-mobile foam and RZ law follow a unique monotonic increasing function (continuous line) that is the extension of the Carman-Kozeny law (equation 3).

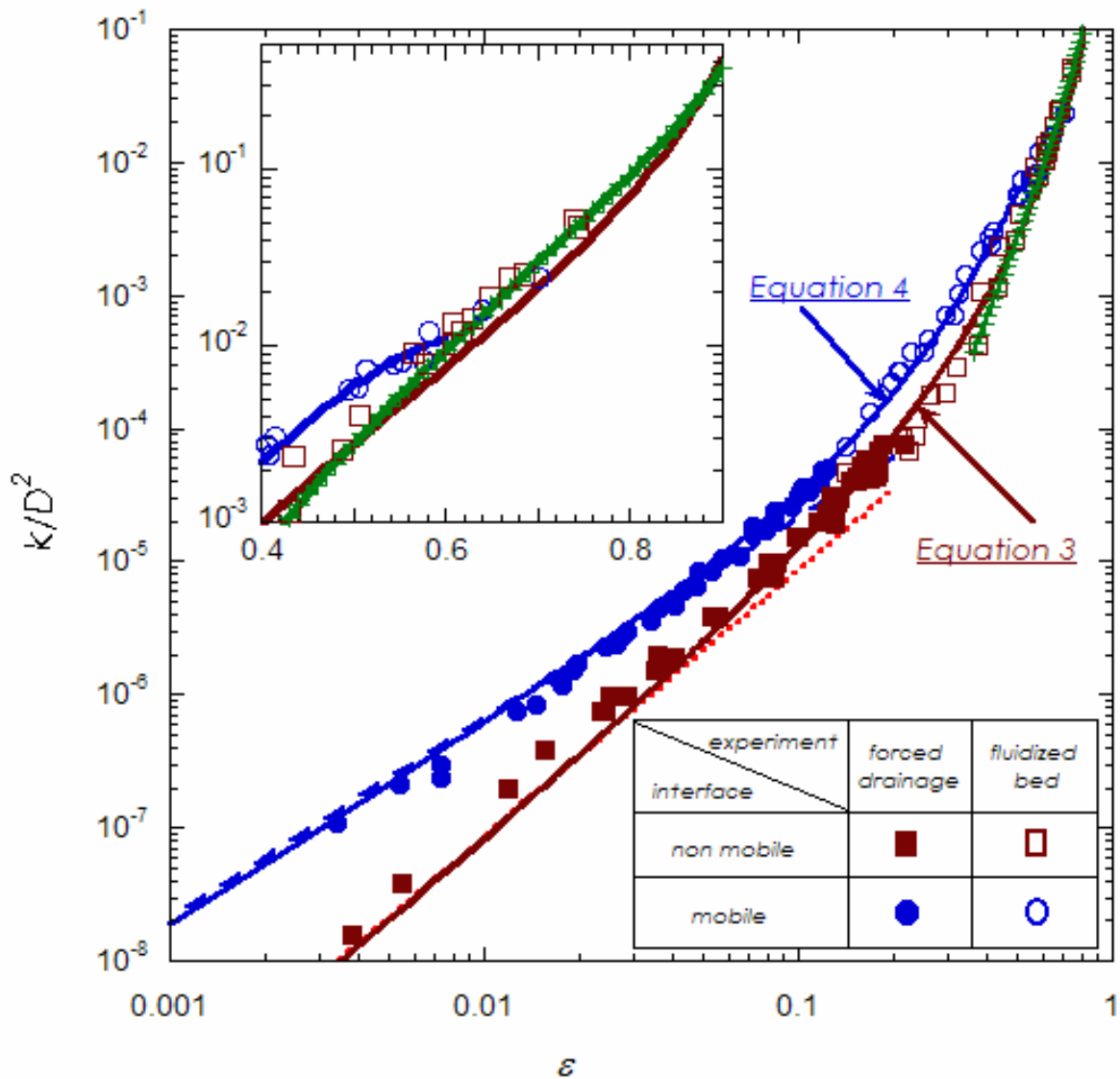

**Figure 4**: Permeability of particulate systems made of monodisperse bubbles as a function of porosity. Models detailed in the text are represented by the red dotted line [7] and the dashed blue line [8]. The Richardson-Zaki law (RZ law) is represented by the green plus sign curve. Experimental data (see legend for details) are compared to the permeability calculated from Carman-Kozeny law (continuous line). Fits proposed in equations 3 and 4 are represented on the graph. Inset : lin-log plot of the permeability as a function of porosity focused on the fluidized domain.

## V. Discussion :

Amazingly, the Carman-Kozeny law (equation 3.1 and 3.2) agrees very well with previous results with non-mobile interfaces over the whole range of porosity explored. This law agrees perfectly with the existing theory for foam made of rigid interfaces at low porosity [7], and also permits us to describe experimental results at higher porosity which is not trivial has recently shown for emulsion by Peron et al. [24]. Moreover, this shows that close to the close packing limit ($\varepsilon$ of the order of 0.3) the normalized specific surface does not vary a lot with porosity – within 10% around 4, whereas permeability varies over more than a decade. We conclude that whatever the structure and the dynamics of the systems (liquid or solid state) the permeability is well predicted at first order by the equations 3.1 and 3.2, meaning that even if the transition from solid to liquid is of great interest, it has little influence on the prediction of the permeability of the sample, if the structure remains random.

To focus on the wet part (extension to the suspension regime), we present in the insert of figure 4 a linear-log plot of the permeability as a function of porosity. The agreement between the data of bubble suspension of non-mobile interfaces, the data of hard sphere suspension (RZ fit) and Carman-Kozeny model is quite good at first order for the range of 0.4-.85 in liquid volume fraction. The maximum relative deviation between the hard sphere suspension (RZ fit) and Carman-Kozeny model is less than 20%. This shows that, at first order, concentrated and semi-dilute suspensions behave like a static porous media even though particles are in permanent agitation due to hydrodynamic interactions [14, 25]. These results are consistent with previous numerical simulations [16], but differs from Davies et al. conclusions, who stated that the constant c of Carman-Kozeny can be much larger than 5 and depend on porosity as well as on particle suspension properties [15]. Here, we demonstrate that the Carman-Kozeny law with the constant c=5 allows to describe relatively well the permeability of suspensions made of monodisperse spherical entities (solid particles or bubbles with non-mobile interface). Note that in agreement with [13] for $Re_p = 4$, a better description of the experimental data at large porosity ($\varepsilon > 0.5$) is obtained by considering $V_{sed} = V_{Stokes}\, \varepsilon^n$ where n is equal to 4.9 in equation 1; the maximum relative deviation is less than 10%.

Moreover, we observe in Figure 4 that at large liquid fractions ($\varepsilon > 0.6$), the permeability of the particulate systems made of mobile interfaces catch up with the Kozeny-Carman law and with the data for non-mobile interfaces. This can be due to a kinetic rigidification of the interfaces of the bubbles made of mobile surfactant. Indeed at this large liquid fractions, Marangoni flows that occur at the surface of disjoined bubbles when the bubbles are made of mobile interface may rigidify the interface [26]. Finally, we propose an experimental fit of the permeability as a function of porosity for an assembly of bubbles made of mobile interfaces:

$$k(\varepsilon) = \frac{\varepsilon^{3/2}}{1700\left(1 - 2.7\varepsilon + 2.2\varepsilon^2\right)^2} D^2 \text{ for } \varepsilon < 0.6 \text{ (equation 4.1)}$$

$$k(\varepsilon) = \frac{\varepsilon^3}{180(1-\varepsilon)^2} D^2 \text{ for } \varepsilon > 0.6 \text{ (equation 4.2)}$$

For convenience, we choose a fit similar to the one proposed for non-mobile interfaces in equation 3. There is not yet a model to predict the mobile case. However, we can guess some physical arguments to incorporate in a future model : i) the lengthscale over which viscous dissipation occurs is smaller than the length scale associated to specific surface area in the Kozeny-Carman model; ii) this length scale varies with porosity.

## VI. Conclusion:

We present experimental data of the permeability of a bubble assembly for porosity varied from 0.15 to 0.75. These new data connect the dry limit and the wet limit already studied for foam and hard-sphere suspension. They permit to show continuous variation of the normalized permeability as function of porosity on a graphic with 7 decades for $k/D^2$ and 3 decades for $\varepsilon$. We propose two sets of functions that fit experimental data obtained for both mobile and non-mobile interfaces. We assume that these functions envelop the permeability curves as a function of porosity for any assembly of monodisperse spherical entities for porosity varied from 0.001 up to 0.85.

For the particular case of non-mobile interfaces, we show that once the dissipation is dominated by viscous forces (low Reynolds number, non-mobile interface), the Carman-Kozeny law and an accurate description of the specific surface allow describing the fluid flow through media made of monodisperse entities for a large range of liquid volume fraction. Indeed, we are able to describe experimental data from 0.001 up to 0.85, with no adjustable parameter (since the parameter c=5 is originally validated on packed bead beds).

We expect that this work is of great interest for other systems, in addition to foam and suspensions, where the continuous liquid phase varies over a wide range, for example emulsions or complex biological systems.

**Acknowledgement:** We thank Pierre Mills and Jerome Martin for fruitful discussions. We thank Jeffrey Morris for valuable comments. We acknowledge the Agence Nationale de la Recherche (ANR-05-JCJC-0234-01), the European Space Agency (MAP "Hydrodynamics of wet foams") and the Centre Nationale d'Etudes Spatiales for fundings.